\documentclass[prx, notitlepage, twocolumn,nofootinbib,superscriptaddress
]{revtex4-1}
\usepackage[colorlinks=true,citecolor=blue,linkcolor=blue]{hyperref}
\usepackage{graphicx}
\usepackage{color}
\usepackage{dcolumn}
\usepackage{bm}
\usepackage{mathrsfs}
\usepackage{amsmath}
\usepackage{amssymb}
\usepackage{amsthm}
\usepackage{amsfonts}
\usepackage{enumerate}
\usepackage{latexsym}
\usepackage{hyperref}
\usepackage{centernot}
\usepackage{multirow}
\usepackage{pgffor}
\usepackage[section]{placeins}
\usepackage{float}

\begin{document}

\title{Towards ideal topological materials: Comprehensive database searches
using symmetry indicators}

\begin{abstract}
Topological materials (TMs) showcase intriguing physical properties defying
expectations based on conventional materials, and hold promise for the
development of devices with new functionalities. While several theoretically
proposed TMs have been experimentally confirmed, extensive experimental
exploration of topological properties as well as applications in realistic
devices have been held back due to the lack of excellent TMs in which
interference from trivial Fermi surface states are minimized. We tackle this
problem in the present work by applying our recently developed method of
symmetry indicators to all non-magnetic compounds in the 230 space groups.
An exhaustive database search reveals {\em thousands} of TM candidates. Of these,
we highlight the excellent TMs, the \textbf{258} topological insulators and
\textbf{165} topological crystalline insulators which have either noticeable
full band gap or a considerable direct gap together with small trivial Fermi
pockets. We also give a list of \textbf{489} topological semimetals with the
band crossing points located near the Fermi level. All predictions obtained
through standard generalized gradient approximation (GGA) calculations were
cross-checked with the modified Becke-Johnson (MBJ) potential calculations,
appropriate for narrow gap materials. With the electronic and optical
behavior around the Fermi level dominated by the topologically non-trivial
bands, these newly found TMs candidates open wide possibilities for
realizing the promise of TMs in next-generation electronic devices.
\end{abstract}

\date{\today }
\author{Feng Tang}
\affiliation{National Laboratory of Solid State Microstructures and School
of Physics, Nanjing University, Nanjing 210093, China} %
\affiliation{Collaborative Innovation Center of Advanced Microstructures,
Nanjing 210093, China}

\author{Hoi Chun Po }
\affiliation{Department of Physics, Harvard University, Cambridge, MA 02138,
USA}

\author{Ashvin Vishwanath}
\affiliation{Department of Physics, Harvard University, Cambridge, MA 02138,
USA}

\author{Xiangang Wan}
\affiliation{National Laboratory of Solid State Microstructures and School
of Physics, Nanjing University, Nanjing 210093, China} %
\affiliation{Collaborative Innovation Center of Advanced Microstructures,
Nanjing 210093, China}
\maketitle

\section{introduction}

Since the discovery of two-dimensional (2D) and three-dimensional (3D)
topological insulators (TIs), band topology in condensed-matter materials
have attracted broad interest \cite{1,2}. Soon after that, the important
role of symmetries in topological materials (TMs) had been recognized, and
various novel topological crystallize phases (TCPs) had been proposed \cite%
{B1,B2,B3}: mirror Chern insulator \cite{3}, hourglass fermions \cite{4},
higher-order topological insulators \cite{5-1,5-2,5-3,5-4,5-5, 5-7},
wallpaper fermions and the nonsymmorphic Dirac insulator \cite{6}, etc. In
addition to the various forms of topological (crystalline) insulators,
topological semimetals (TSMs) \cite{7} with distinct quasi-particle
properties, such as Dirac \cite{8,9,10}, Weyl \cite{A1,12,13}, nodal-line
\cite{14}, nodal-chain \cite{14-2}, and those from three-fold (or higher)
band degeneracies \cite{15}, etc, had also attracted a lot of attention.
Exploring their novel properties and their connections to fundamental
physics is one of the central research topics of contemporary
condensed-matter physics and materials science.

TMs could be characterized by their robust, unconventional properties \cite%
{1,2,7}: like the presence of surface/edge states, absence of
backscattering, topological Fermi arc surface states, highly anisotropic
negative longitudinal magnetoresistance from the chiral anomaly, nontrivial
quantum oscillations, nonlocal transport bebaviors, photoinduced anomalous
Hall effects, non-Abelian statistics due to the spin momentum locked and the
associated superconductivity, and many other novel quantum behavior.
Leveraging these properties, TMs could enable low-power and high-speed
devices for applications in advanced spintronic, magnetoelectric, and
optoelectronic industry. However, despite intense research efforts,
excellent TM candidates for realistic technological application remain
difficult to come by.

As one of the best known TIs, Bi$_{2}$Se$_{3}$ family has bulk band gaps of
typically a few 100 meV \cite{Bi2Se3-1,Bi2Se3-2}. Unfortunately, native
defects tend to always dope their chemical potential into the bulk
conduction or valence band \cite{Bi2Se3-3}, which leads to metallic states
that mask the topological response of the otherwise insulating bulk. Only
though recent heroic efforts tailored at modifying the chemistry have better
insulating properties been achieved \cite{Cava}. Existing TCP materials \cite%
{B3} also suffer from similar problems. For instance, the mirror-Chern
insulator SnTe \cite{3} is always heavily \textit{p}-type and shows unwanted
bulk conductivity \cite{SnTe-no-good}.

As for Weyl semimetals like TaAs, the Weyl points are usually located
slightly away from the Fermi level and the node separation in momentum space
is typically small \cite{12,13}, this leads to the coexistence of trivial
Fermi pockets and small Fermi arcs. On the other hand, the Dirac nodes in Na$%
_{3}$Bi\cite{9} and Cd$_{3}$As$_{2}$\cite{10}\ are indeed exactly located at
Fermi level. However, the former is not air-stable and the latter is toxic,
which limits their potential for extensive experimental studies and
technological applications.

Very recently, by integrating recent advances in the theory of TCPs \cite%
{Po-1,QCT,Magnetic}, in particular the theory of symmetry-based indicators
(SIs) developed in Ref.\ \cite{Po-1,Magnetic}, into \textit{ab initio}
calculations \cite{Tang-1,Tang-2}, we have developed a theoretical method
for systematically discovering TMs \cite{Tang-1}. Our algorithm efficiently
categorizes any given material into one of the three cases \cite{Tang-1}:
\textbf{Case 1}: possibly an atomic insulator; \textbf{Case 2}: definitely
topologically nontrivial; \textbf{Case 3}: definitely owing symmetry
protected band crossings. We predicted a large number of TCPs based on this
method in Refs.\ \cite{Tang-1,Tang-2}, and some of our predictions had
already been experimentally confirmed \cite{MoTe2,Yao}. In this work, we
perform study on all the suitable non-magnetic materials presented in
structural databases \cite{str-database}.

\section{Materials search}

We employ the paradigm of materials search detailed in Ref.\ %
\onlinecite{Tang-1}, which can be summarized into three steps for any space
group. First, we compute a basis for the space of symmetry representations
through an inspection of the atomic insulators (please refer to the
supplementary materials (SM)\cite{SM} for a tabulation of the basis found).
Second, based on \textit{ab initio} calculations we compute the irreducible
representations for the electronic bands at the high-symmetry momenta for
all the non-magnetic stoichiometric materials listed on the Inorganic
Crystal Structure Database (ICSD)\textbf{\ }\cite{str-database}. Third, we apply the theory of symmetry indicators \cite{Po-1} to expose the
TM candidates.

While there are more than 180000 crystal structures listed in the ICSD \cite%
{str-database}, the vast majority of them are non-stoichiometric and/or
contain typical magnetic ions. For most of materials with typical magnetic
ions in the structural databases \cite{str-database}, the experimental
feature, magnetic or nonmagnetic, is unavailable. Moreover, electronic
correlations typically have pronounced effects on the electronic and
magnetic properties of compounds containing 3$d$/4$f$ ions, and one usually need
LDA+U or LDA+DMFT scheme, which have adjustable parameter U, to take into
account for the Coulomb interactions in these systems \cite{DMFT}. To
improve the reliability of our predictions, we thus remove such systems from
our materials search. Similarly, we also remove 5$f$ compounds, with the added
concern that the majority of them are radioactive and are unsuitable for
realistic applications. These filters bring the number of relevant materials
from all the 230 space groups down to 17258, which we pass on to the next
step of the analysis. To improve the reliability of our materials
prediction, we crosscheck results obtained from the standard generalized
gradient approximation (GGA)\cite{PBE} calculation against that from
modified Becke-Johnson (MBJ) potential-based calculations \cite{MBJ}
(Methods). This minimizes the possibility of misdiagnosing the topological
properties of materials arising from the systematic underestimation of band
gaps in GGA \cite{Yao-MBE}.

As in our previous works \cite{Tang-1,Tang-2}, we find a good fraction of
the materials (\textbf{11269 in 17258})\ to be topological, although most of
them have dirty Fermi surfaces arising from coexisting trivial states.
Instead of listing all the TMs we found, we focus on the nearly ideal
candidates with clean Fermi surfaces. In the main text, we discuss some
representatives TMs predicted from both GGA and MBJ calculations; in the SM
\cite{SM} we tabulate all the other nearly ideal candidates found. We remark
that approximately $25\%$ of the TMs identified from GGA are found to be
trivial from MBJ calculations. This highlights the importance of
crosschecking the two methods, and we indicate their (dis-)agreements in the
tables of the SM\cite{SM}. We also give the atomic insulator basis for each
space group ($\mathcal{SG}$), a key to our theoretical method, in the SM {%
\cite{SM}}, so that one can easily use our scheme for diagnosing the TMs.

\begin{figure}[tbp]
\caption{The MBJ calculated electronic band structures of the representative
strong TIs discovered in our search: (a) Ag$_{2}$Zr ($\mathcal{SG}\mathbf{139%
}$); (b) Ba$_{11}$Bi$_{14}$Cd$_{8}$ ($\mathcal{SG}\mathbf{12}$).}%
\includegraphics[width=8cm]{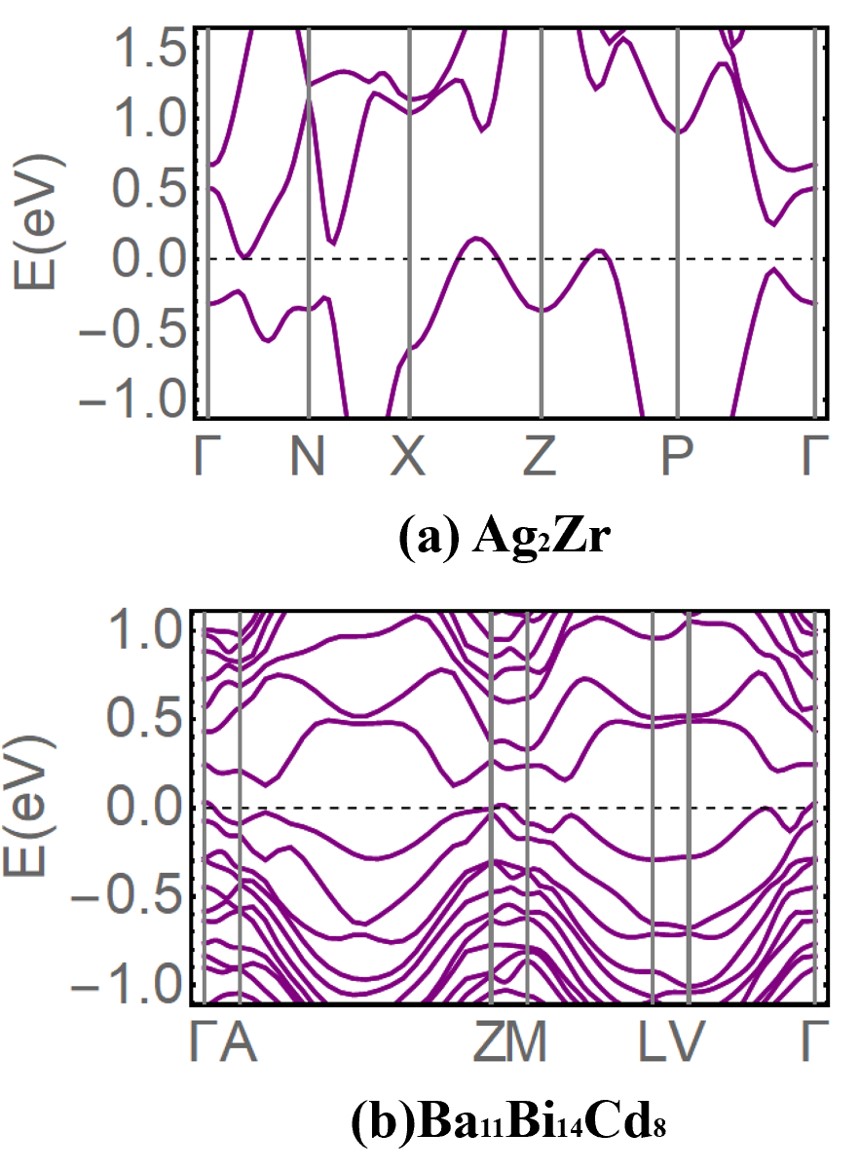}\newline
\label{TI}
\end{figure}

\section{Topological insulators}

We list all the identified strong topological insulators in Table I of the
SM \cite{SM}, and we show two\textbf{\ }representative materials, one with a
full gap and the other has large direct gaps. We analyze them in details as
follows. Ag$_{2}$Zr crystalizes in the body-centered orthorhombic Bravais
lattice in the $\mathcal{SG\ }$ \textbf{139}. The SI group is $X_{\mathrm{BS}%
}=\mathbb{Z}_{2}\times \mathbb{Z}_{8}$ \cite{Po-1}. The 46 valence bands are
found to take an SI of $(0,1)$, indicating it is a strong TI which could
also be diagnosed using the Fu-Kane parity criterion \cite{Fu-Kane}. As
shown in Fig. \ref{TI}(a), there are continuous large direct gaps ($\sim 300$
meV) throughout the whole BZ, and the small pockets just contributes a
little density of states at the Fermi level.

Our second strong TI candidate is Ba$_{11}$Bi$_{14}$Cd$_{8}$, which
crystalizes in space group $\mathbf{12}$. The SI group is $X_{\mathrm{BS}}=%
\mathbb{Z}_{2}\times \mathbb{Z}_{2}\times \mathbb{Z}_{4}$ \cite{Po-1}, and
we found the SI to be $(0,1,1)$, indicating that it is a strong TI. It is
worth mentioning that Ba$_{11}$Bi$_{14}$Cd$_{8}$ has a full band gap as is
shown in the MBJ band plot in Fig. \ref{TI}(b). And its band gap ($\sim 34\
meV$ by MBJ calculations) is larger than  the room temperature. Thus, it is
interesting to see if Ba$_{11}$Bi$_{14}$Cd$_{8}$ can display robust
topological features.
\begin{figure}[tbp]
\caption{The MBJ calculated electronic band structures of the representative
TCIs discovered in our search: (a) BaGe ($\mathcal{SG}\mathbf{63}$); (b) Bi$%
_{2}$Se$_{2}$ ($\mathcal{SG}\mathbf{164}$).}
\label{TCI}\includegraphics[width=7cm]{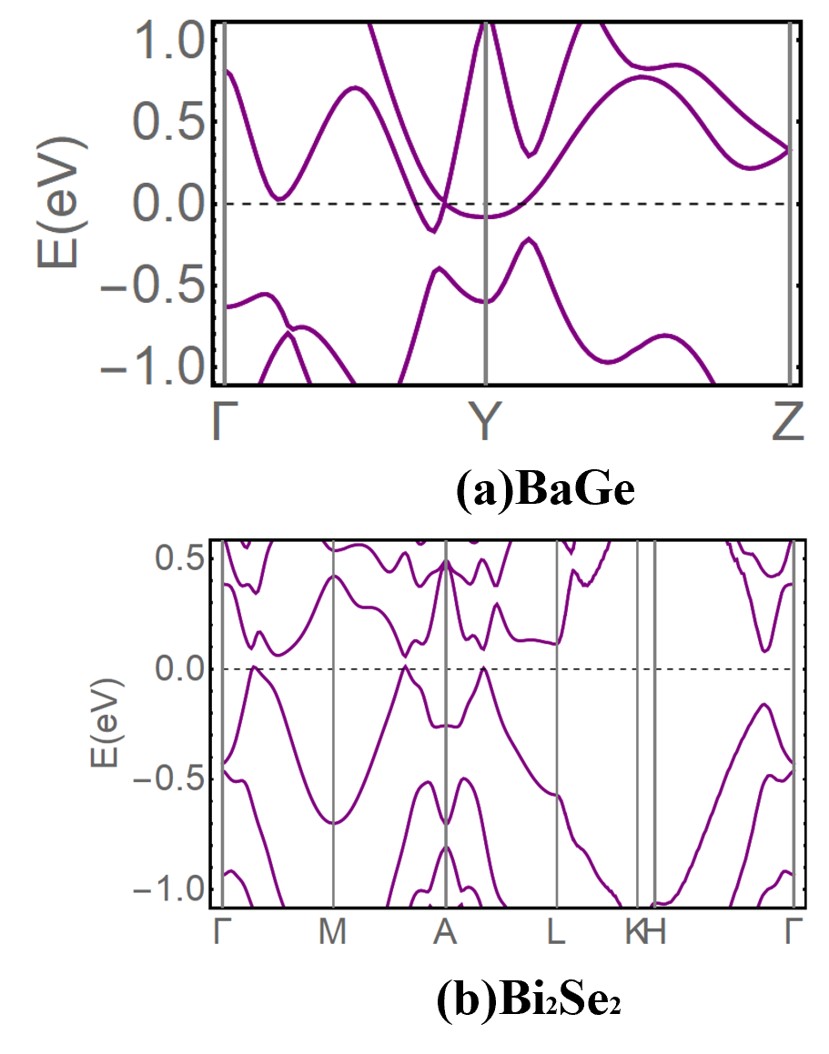}\newline
\end{figure}

\section{Topological crystalline insulators}

While time-reversal symmetry protects the strong TI, the richness of spatial
symmetries leads to a great variety of topological crystalline phases \cite%
{B1,B2,B3,Po-1,QCT,Magnetic,5-7,TCP-1,TCP-2,TCP-3}. From the many materials
candidates for topological crystalline insulators we discovered, we choose
two representatives to discuss in details: BaGe (Fig. \ref{TCI}(a)) with
sizable direct gap ($\sim 300$ meV) and Bi$_2$Se$_2$ (Fig. \ref{TCI}(b))
with a full bulk gap.

BaGe crystalizes in the space group $\mathbf{63}$, whose SI group is $X_{%
\mathrm{BS}}=\mathbb{Z}_{2}\times \mathbb{Z}_{4}$ \cite{Po-1,SM}. It is
known that if the strong SI in $\mathbb{Z}_{4}$ is odd, i.e., $1$ or $3$,
the material must be a strong TI. However, in general the SI does not fully
map out the band topology: when the value is even one only knows that the
strong index is trivial, and the precise TCP it is in can only be determined
through further analysis \cite{5-7,TCP-2}. Here, we found the SI $(1,0)\in
\mathbb{Z}_{2}\times \mathbb{Z}_{4}$, which indicates that BaGe is not a
strong TI and must be a weak TI. Due to the described ambiguity, however,
further analysis, like the computation of mirror Chern numbers, must be
supplied to ascertain the precise phase of BaGe.

Next, we consider Bi$_{2}$Se$_{2}$ in $\mathcal{SG}$ $\mathbf{164}$. For
this $\mathcal{SG}$ the SI group was found to be $X_{\mathrm{BS}}=\mathbb{Z}%
_{2}\times \mathbb{Z}_{4}$ \cite{Po-1,SM}. We found that the SI of Bi$_{2}$Se%
$_{2}$ is $(1,2)$, which implies it must be a TCI. For the \textit{ab initio}
calculated parities, we find that the 3D topological invariant $(\nu _{0};\nu _{1},\nu
_{2},\nu _{3})=(0;1,1,1)$, indicating that it is a weak TI. {Furthermore,
from the results of Refs.\ \onlinecite{TCP-2,5-7} we know that it either has
mirror-protected surface Dirac cones, or feature rotation-protected Dirac
cones together with coexisting hinge states. }

\begin{figure}[tbp]
\includegraphics[width=8cm]{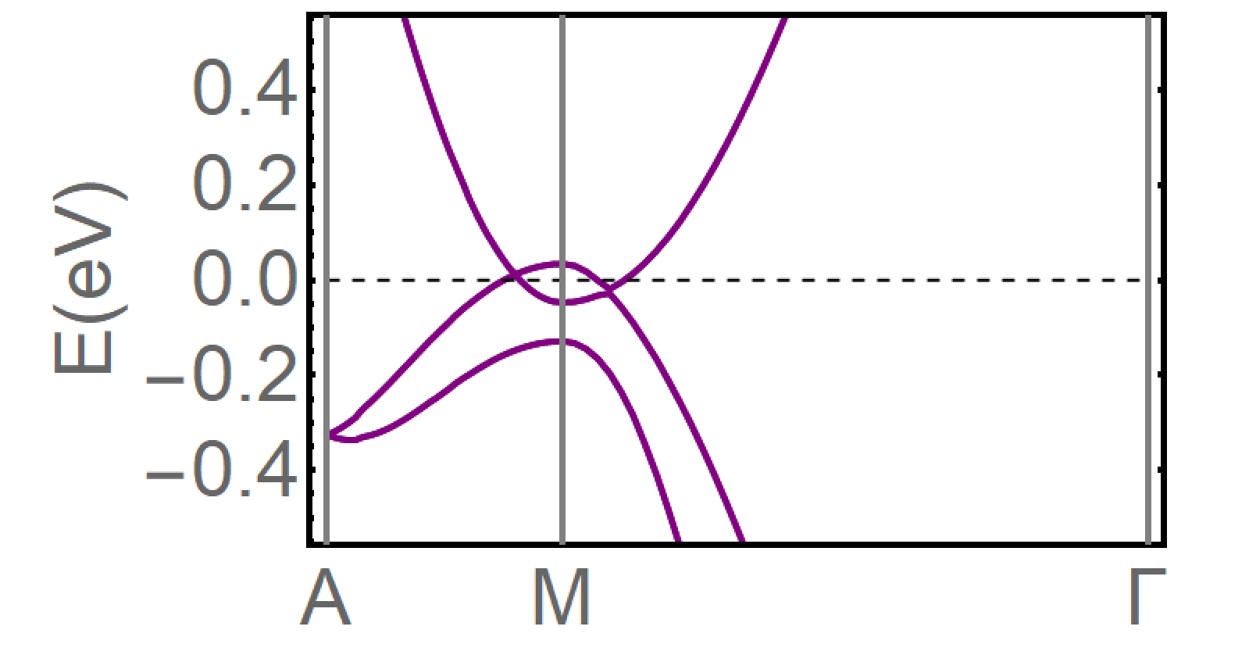}\newline
\caption{The MBJ calcualted electronic band structures of the representative
nearly ideal Dirac semimetal discovered in our search: PdO ($\mathcal{SG}%
\mathbf{131}$).}
\label{DSM}
\end{figure}

\section{Topological Dirac semimetal}

We also discovered a lot of topological semimetals and give a representative
Dirac semimetal as follows: PdO in $\mathcal{SG}\mathbf{131}$. PdO
crystallizes in the tetragonal structure whose point group is $D_{4h}$. Our
algorithm indicates that the material belong to \textbf{Case 3} \cite{Tang-1}%
, which have necessary band crossings near the Fermi energy. We then find
that such a band crossing happens along the high-symmetry line $AM$ (Fig. %
\ref{DSM}), resulting in a Dirac point. This Dirac point is protected by the
$C_{4v}$ symmetry, and is stable unless the strain or pressure lowers the
symmetry gapping the Dirac point. The Fermi level nearly intersects the band
crossing point, thus the low energy physics is governed mainly by the Dirac
physics, making it a promising material for applications.

\section{Discussion}

The predictions of the topological properties of the above materials \emph{%
are based on not only the GGA but also the MBJ method}. While we discuss
several representative TMs in main text, we also tabulate all the other
nearly ideal TMs candidates we found using the GGA method in the SM \cite{SM}%
, alongside with the result of our MBJ cross-check on the topological
properties of the TIs and TCIs. Importantly, the stringent criteria we
applied bring out the 8\% of nearly ideal TM candidates from the thousands
found in the search, which could expedite experimental progress. Among the
huge list of TM candidates we discovered, some of the big-gap systems may be
nontoxic and are free of native defects, which enables the growth of
high-quality samples. This would hopefully help overcome the limitations of
previously proposed candidates and move TMs towards applications.

\section{Methods}

Our \textit{ab initio} calculations are based on WIEN2K, one of the most
accurate packages \cite{W2K}. We consider spin-orbital coupling in all of
our calculations. The standard generalized gradient approximation (GGA) with
Perdew-Burke-Ernzerhof (PBE) realization was adopted for the
exchange-correlation functional \cite{PBE}. However, it is well known that
GGA calculation tends to underestimate the band gap, an important aspect for
TMs, and consequently the GGA-based prediction of TMs is not always reliable
\cite{Yao-MBE}. In contrast, the modified Becke-Johnson (MBJ) potential\cite%
{MBJ} gives band gaps which are in good agreement with experiments for a
wide variety of simple semiconductors and insulators \cite{MBJ,MBJ-1,MBJ-2}.
Moreover, it also produces accurate prediction on band ordering, which is
key to the determination of band inversion and hence the topological
properties of materials \cite{Yao-MBE}. For example, LaSb was identified by
GGA to be a topological insulator, but experimentally it was found to be
topological trivial, in line with the MBJ prediction \cite{MBJ-LaSb}. Thus,
we further cross-check the GGA results by MBJ calculations \cite{MBJ}.

\begin{acknowledgements}
FT and XGW were supported by National Key R\&D Program of China (No.\ 2017YFA0303203 and 2018YFA0305700), the NSFC (No.\ 11525417, 51721001 and 11790311). FT was also supported by the program B for Outstanding PhD candidate of Nanjing University. XGW was partially supported by a QuantEmX award funded by the Gordon and Betty Moore Foundation's EPIQS Initiative through ICAM-I2CAM, Grant GBMF5305 and by the  Institute of Complex Adaptive Matter (ICAM).
AV is supported by NSF DMR-1411343, and by a Simons Investigator Grant.
\end{acknowledgements}

\newpage
\clearpage

\onecolumngrid
\setcounter{page}{0}

\foreach \t in {1,2,3,4,5,6,7,8,9,10,11,12,13,14,15,16,17,18}
{

\begin{figure*}
\centering
  \includegraphics[width=20cm]{081314152264_\t.pdf}
\end{figure*}
}
\clearpage
\foreach \t in {19,20,21,22,23,24,25,26}
{

\begin{figure*}
\centering
  \includegraphics[width=20cm]{081314152264_\t.pdf}
\end{figure*}
}
\clearpage
\foreach \t in {27,28,29,30,31,32,33,34}
{

\begin{figure*}
\centering
  \includegraphics[width=20cm]{081314152264_\t.pdf}
\end{figure*}
}
\clearpage
\foreach \t in {35,36,37,38,39,40,41,42}
{

\begin{figure*}
\centering
  \includegraphics[width=20cm]{081314152264_\t.pdf}
\end{figure*}
}
\clearpage
\foreach \t in {43,44,45,46,47,48,49,50}
{

\begin{figure*}
\centering
  \includegraphics[width=20cm]{081314152264_\t.pdf}
\end{figure*}
}
\clearpage
\foreach \t in {51,52,53,54,55,56,57,58}
{

\begin{figure*}
\centering
  \includegraphics[width=20cm]{081314152264_\t.pdf}
\end{figure*}
}
\clearpage
\foreach \t in {59,60,61,62,63,64,65,66}
{

\begin{figure*}
\centering
  \includegraphics[width=20cm]{081314152264_\t.pdf}
\end{figure*}
}
\clearpage
\foreach \t in {67,68,69,70,71,72,73,74}
{

\begin{figure*}
\centering
  \includegraphics[width=20cm]{081314152264_\t.pdf}
\end{figure*}
}
\clearpage
\foreach \t in {75,76,77,78,79,80,81,82}
{

\begin{figure*}
\centering
  \includegraphics[width=20cm]{081314152264_\t.pdf}
\end{figure*}
}
\clearpage
\foreach \t in {83,84,85,86,87,88,89,90}
{

\begin{figure*}
\centering
  \includegraphics[width=20cm]{081314152264_\t.pdf}
\end{figure*}
}
\clearpage
\foreach \t in {91,92,93,94,95,96,97,98}
{

\begin{figure*}
\centering
  \includegraphics[width=20cm]{081314152264_\t.pdf}
\end{figure*}
}
\clearpage
\foreach \t in {99,100,101,102,103,104,105,106}
{

\begin{figure*}
\centering
  \includegraphics[width=20cm]{081314152264_\t.pdf}
\end{figure*}
}
\clearpage
\foreach \t in {107,108,109,110,111,112,113,114}
{

\begin{figure*}
\centering
  \includegraphics[width=20cm]{081314152264_\t.pdf}
\end{figure*}
}
\clearpage
\foreach \t in {115,116,117,118,119,120,121,122}
{

\begin{figure*}
\centering
  \includegraphics[width=20cm]{081314152264_\t.pdf}
\end{figure*}
}
\clearpage
\foreach \t in {123,124,125,126,127,128,129,130}
{

\begin{figure*}
\centering
  \includegraphics[width=20cm]{081314152264_\t.pdf}
\end{figure*}
}
\clearpage
\foreach \t in {131,132,133,134,135,136,137,138}
{

\begin{figure*}
\centering
  \includegraphics[width=20cm]{081314152264_\t.pdf}
\end{figure*}
}
\clearpage
\foreach \t in {139,140,141,142,143,144,145,146}
{

\begin{figure*}
\centering
  \includegraphics[width=20cm]{081314152264_\t.pdf}
\end{figure*}
}
\clearpage
\foreach \t in {147,148,149,150,151,152,153,154}
{

\begin{figure*}
\centering
  \includegraphics[width=20cm]{081314152264_\t.pdf}
\end{figure*}
}
\clearpage
\foreach \t in {155,156,157,158,159}
{

\begin{figure*}
\centering
  \includegraphics[width=20cm]{081314152264_\t.pdf}
\end{figure*}
}
\clearpage
\end{document}